\documentclass[aps,pre
,showpacs
,twocolumn
,superscriptaddress
]
{revtex4-1}
\pdfoutput=1
\usepackage{amsmath,amsthm,amssymb}
\usepackage{array}
\usepackage{graphicx}
\usepackage{fancyhdr}
\usepackage{color}
\usepackage{extarrows}
\usepackage{enumerate}
\usepackage{epstopdf}
\usepackage[colorlinks,
          linkcolor=black,
            citecolor=black,
            urlcolor=blue
           ]{hyperref}
\usepackage{natbib}

\begin{document}

\title{Non-equilibrium work relation beyond Boltzmann-Gibbs distribution}

\author{Ying Tang}
\email{Corresponding author. Email: jamestang23@gmail.com}
\affiliation{Department of Physics and Astronomy, Shanghai Jiao Tong University, Shanghai 200240, China}
\affiliation{Key Laboratory of Systems Biomedicine Ministry of Education, Shanghai Center for Systems Biomedicine, Shanghai Jiao Tong University, Shanghai 200240,
China}
\author{Ruoshi Yuan}
\affiliation{School of Biomedical Engineering, Shanghai Jiao Tong University, Shanghai 200240, China}
\author{Ping Ao}
\email{Corresponding author. Email: aoping@sjtu.edu.cn}
\affiliation{Key Laboratory of Systems Biomedicine Ministry of Education, Shanghai Center for Systems Biomedicine, Shanghai Jiao Tong University, Shanghai 200240,
China}
\affiliation{Department of Physics and Astronomy, Shanghai Jiao Tong University, Shanghai 200240, China}

\date{\today}

\begin{abstract}
The presence of multiplicative noise can alter measurements of forces acting on nanoscopic objects. Taking into account of multiplicative noise, we derive a series of non-equilibrium thermodynamical equalities as generalization of the Jarzynski equality, the detailed fluctuation theorem and the Hatano-Sasa's relation. Our result demonstrates that the Jarzynski equality and the detailed fluctuation theorem remains valid only for systems with the Boltzmann-Gibbs distribution at the equilibrium state, but the Hatano-Sasa's relation is robust with respect to the ambiguity on multiplicative noise.
\end{abstract}
\maketitle

\section{Introduction}
\label{section 1}
The calculation on free energy changes is a central endeavor of non-equilibrium physics. A series of remarkable equalities, such as the Jarzynski equality \cite{jarzynski1997nonequilibrium}, the Hatano-Sasa's relation \cite{PhysRevLett.86.3463}, and the fluctuation theorem \cite{crooks1999entropy,Kim2001Fluctuation,PhysRevLett.89.050601,kurchan2007non,ge2008generalized,PhysRevLett.100.250601,seifert2012stochastic},
enable the calculation on free energy changes from repeated non-equilibrium force measurements \cite{Liphardt07062002,Trepagnier19102004,Hummer14122010}. However, a recent experiment for a Brownian particle near a wall demonstrates that multiplicative noise alters measurements of forces on nanoscopic objects \cite{volpe2010influence}. The force-measurement process thus requires multiplicative noise to be carefully taken into account. The uncertainty caused by multiplicative noise can be traced to the controversy on choosing the interpretation for stochastic dynamics \cite{gardiner2004handbook,ao2007existence}. Therefore, how this uncertainty may affect the calculation of free energy changes from non-equilibrium force measurements is crucial to be explored.

In this paper, we provide a series of non-equilibrium thermodynamical equalities compatible for the general stochastic interpretation on multiplicative noise. They can be regarded as generalization of the Jarzynski equality, the detailed fluctuation theorem and the Hatano-Sasa's relation. Our result Eq.~(\ref{generalized JE}) demonstrates that repeated non-equilibrium work measurements with the Jarzynski equality does not lead the free energy change. Instead, it corresponds to a ratio of generalized partition functions at the initial and final equilibrium states. This generalized partition function can quantify the effect of multiplicative noise and corresponds to the Helmholtz free energy in the case of the Boltzmann-Gibbs distribution as the equilibrium state distribution, where anti-Ito's interpretation is preferable \cite{hanggi1978derivations,ao2007existence}.

The generalized fluctuation theorem Eq.~(\ref{generalized FT}) can also reduce to the conventional detailed fluctuation theorem for the non-equilibrium work distribution under anti-Ito's interpretation. Interestingly, the generalized Hatano-Sasa's relation Eq.~(\ref{generalized HS}) has the same form as the conventional one even under the general stochastic interpretation. This demonstrates that the Hatano-Sasa's relation is robust with respect to the uncertainty on multiplicative noise.

This paper is organized as follows. In Sec.~\ref{section 2}, we provide the generalization of the Jarzynski equality, the detailed fluctuation theorem, and the Hatano-Sasa's relation. In Sec.~\ref{section 3}, we given the detailed derivation on our main result by introducing the reverse process and the path integral framework. In Sec.~\ref{section 4}, we summarize our work. In the Appendix, we give the construction on the path integral formulation from the overdamped Langevin dynamics.

\section{Generalized non-equilibrium work relations}
\label{section 2}

For the convenience of comparison, we first state the Jarzynski equality:
\begin{align}
\label{Jarzynski}
\langle \exp(-W)\rangle=\exp(-\Delta \mathcal{F})=\frac{Z_{1}[\lambda(t_{N})]}{Z_{1}[\lambda(t_{0})]},
\end{align}
where $\langle\cdots\rangle$ denotes an ensemble average of measurements on work, $\Delta \mathcal{F}$ is the Helmholtz free energy change between two equilibriums, and $Z_{1}[\lambda(t_{N})]$ is the partition function corresponding to the Helmholtz free energy. The Boltzmann constant multiplying the temperature ($1/\beta=k_{B}T$) is set to be a unit. In this equation, $W$ denotes values of work defined as: $W^{F}\doteq\int_{t_{0}}^{t_{N}}dt\frac{\partial H}{\partial\lambda}\dot{\lambda}^{F}$ \cite{jarzynski1997nonequilibrium}. The system evolves as $\lambda(t)$ changes according to a protocol, $\lambda^{F}(t)$ from time $t_{0}$ to $t_{N}$, where the superscript $F$ means the forward process. The system is assumed initially in a equilibrium state distribution (e.g. Boltzmann-Gibbs distribution) with $\lambda(t_{0})$, and converges to the same form of equilibrium state distribution with $\lambda(t_{N})$ after the completion of the manipulation.

To discover the effect of multiplicative noise on the non-equilibrium work relations, we consider the overdamped Langevin equation with multiplicative noise in one dimension as our model:
\begin{align}
\label{Langevin1}
\dot{x}=f(x;\lambda)+g(x;\lambda)\xi(t),
\end{align}
where $x$ denotes the position, $\dot{x}$ denotes its time derivative, $f(x;\lambda)$ is the drift term, and $g^{2}(x;\lambda)/2$ is the diffusion coefficient. The parameter $\lambda$ can denote a set of control parameters reflecting such as the influence of an external force to the system \cite{jarzynski1997nonequilibrium}. Here, $\xi(t)$ is a Gaussian white noise with  $\langle\xi(t)\rangle=0$, $\langle\xi(t)\xi(s)\rangle=\delta(t-s)$, where the average is taken with respect to the noise distribution. We study the natural boundary condition in this paper.

For this Langevin equation, an ambiguity in choosing the integration method leads to different stochastic interpretations \cite{gardiner2004handbook}, and a general notation is the so-called $\alpha$-interpretation \cite{shi2012relation}: the values $0$, $1/2$, $1$ of $\alpha$ correspond to Ito's, Stratonovich's \cite{gardiner2004handbook}, and anti-Ito's \cite{hanggi1978derivations} separately. Then, for the Langevin equation under the $\alpha$-interpretation, the corresponding dynamical process for the probability distribution is given by the following Fokker-Planck equation \cite{shi2012relation}:
\begin{align}
\label{Fokker-Planck}
\partial_{t}\rho(x,t)&=-\partial_{x}\Big[\Big(f+\alpha g^{'}g\Big)\rho(x,t)\Big]+\frac{1}{2}\partial^{2}_x\Big[g^{2}\rho(x,t)\Big],
\end{align}
where the superscript prime denotes derivative to $x$. We assume that the probability distribution described by the Fokker-Planck equation converges to  the normalized distribution:
\begin{align}
\rho_{eq}(x;\lambda)=\frac{1}{Z_{\alpha}(\lambda)}\exp[-V_{eq}(x;\lambda)],
\end{align}
where the generalized partition function is:
\begin{align}
Z_{\alpha}(\lambda)\doteq\int^{+\infty}_{-\infty}\exp[-V_{eq}(x;\lambda)]dx.
\end{align}
By the condition that the probability current in the Fokker-Planck equation is vanishing under natural boundary condition for the equilibrium state:
$j(x,t)\doteq[f+(\alpha-1)g^{'}g]\rho(x,t)-(g^{2}/2)\partial_{x}\rho(x,t)=0$, the explicit form for the ``equilibrium state potential'' is \cite{ao2007existence}:
\begin{align}
\label{equilibrium potential}
V_{eq}(x;\lambda)=\phi(x;\lambda)+(1-\alpha)\ln g^{2}(x;\lambda).
\end{align}

Here, $\phi(x;\lambda)$ denotes the potential function constructed in the overdamped Langevin dynamics Eq.~(\ref{Langevin1}). It satisfies  $f(x;\lambda)=-(g^{2}(x;\lambda)/2)\partial_{x}\phi(x;\lambda)$ with the help of the Einstein relation \cite{ao2004potential,volpe2010influence,arenas2012hidden}. It corresponds to the Boltzmann-Gibbs distribution at the equilibrium state: $\rho_{BG}(x;\lambda)=\exp[\mathcal{F}(\lambda)-\phi(x;\lambda)]$, where $\mathcal{F}(\lambda)$ is the Helmholtz free energy for the canonical ensemble. For the system without detailed balance, this potential function can also be constructed with the aid of the generalized Einstein relation \cite{kwon2005structure,ao2008emerging}. From the view of dynamical system, the potential function $\phi$ serves as the Lyapunov function guiding dynamics to the attractor \cite{PhysRevE.87.012708,PhysRevE.87.062109,Yuan2014Lyapunov}, and equals to the Hamiltonian with symplectic structure in a limiting case \cite{ao2004potential}.

Taking into account of multiplicative noise, we obtain the generalized the Jarzynski equality for the overdamped Langevin dynamics:
\begin{align}
\label{generalized JE}
\langle \exp(-\widetilde{W})\rangle=\frac{Z_{\alpha}[\lambda(t_{N})]}{Z_{\alpha}[\lambda(t_{0})]},
\end{align}
where $\widetilde{W}$ denotes values for the work along a single trajectory defined as:
\begin{align}
\label{work1}
\widetilde{W}^{F}\doteq\int_{t_{0}}^{t_{N}}\dot{\lambda}^{F}\frac{\partial{V_{eq}}}{\partial\lambda}dt.
\end{align}
The work defined by Eq.~(\ref{work1}) is the external work done on the system \cite{ge2008generalized}. It is physically measurable in experiments, and repeated measurements on it for the non-equilibrium process leads to the free energy difference between equilibriums \cite{hummer2001free,Trepagnier19102004}. An interesting special case is that when the diffusion coefficient is independent of the control parameter, i.e. when $g(x,\lambda)$ becomes $g(x)$. In this situation, the work $\widetilde{W}^{F}$ is equivalent to $\int_{t_{0}}^{t_{N}}\dot{\lambda}^{F}(\partial\phi/\partial\lambda)dt$ \cite{ao2008emerging}, and $\widetilde{W}=W$ for the system governed by Hamiltonian even under the general stochastic interpretation.

Under anti-Ito's interpretation ($\alpha=1$), $Z_{\alpha}(\lambda)$ becomes the conventional partition function corresponding to the Helmholtz free energy: $\mathcal{F}(\lambda)=-\ln Z_{1}(\lambda)$ \cite{ao2008emerging}. The value $\alpha=1$ is also verified as the proper choice for calculations on the mesoscopic forces in a recent experiment \cite{volpe2010influence}. The reason is that anti-Ito's interpretation leads to the Boltzmann-Gibbs distribution at the equilibrium state \cite{ao2007existence}, i.e. $\rho_{eq}(x;\lambda)=\rho_{BG}(x;\lambda)$. In this case, Eq.~(\ref{generalized JE}) reduces to Eq.~(\ref{Jarzynski}). As a result, Eq.~(\ref{generalized JE}) demonstrates that the free energy difference could be accurately calculated by non-equilibrium work measurements using the Jarzynski equality only for the system with the Boltzmann-Gibbs distribution at the equilibrium state. For general cases, $\ln\langle \exp(-\widetilde{W})\rangle$ does not corresponds to the Helmholtz free energy difference between two equilibriums.

The above result Eq.~(\ref{generalized JE}) is obtained from the generalized fluctuation theorem:
\begin{align}
\label{generalized FT}
\frac{\rho^{F}(\widetilde{W})}{\rho^{R}(-\widetilde{W})}&=\frac{Z_{\alpha}[\lambda(t_{N})]}{Z_{\alpha}[\lambda(t_{0})]}\exp(\widetilde{W}),
\end{align}
with the detailed derivation through the path integral framework in the following. Similarly, only when $\alpha=1$, Eq.~(\ref{generalized FT}) can reduce to the detailed fluctuation theorem for the non-equilibrium work distribution \cite{crooks1999entropy}:
\begin{align}
\label{fluctuation theorem}
\frac{\rho^{F}(W)}{\rho^{R}(-W)}&=\exp(W-\Delta \mathcal{F}).
\end{align}

The generalized Jarzynski equality describes the relation between the free energy change and the work. Another equality about the equilibrium state distribution $\rho_{eq}$ is given by the generalized Hatano-Sasa's relation. To see this, we can rewrite Eq~(\ref{generalized JE}) as:
\begin{align}
\label{generalized HS}
\Big\langle\exp\Big[\int_{t_{0}}^{t_{N}}dt\frac{\partial\ln\rho_{eq}(x;\lambda)}{\partial\lambda}\dot{\lambda}^{F}\Big]\Big\rangle=1.
\end{align}
Note that Eq.~(\ref{generalized HS}) is the same as the conventional Hatano-Sasa's relation \cite{PhysRevLett.86.3463,Trepagnier19102004} even for the general stochastic interpretation. Thus, Eq.~(\ref{generalized HS}) shows that the Hatano-Sasa's relation is robust with respect to the stochastic interpretation.

The previous work on the non-equilibrium work relations for systems with multiplicative noise \cite{chernyak2006path} is derived by the path integral formulation under the Stratonovich's interpretation. Their work is different from our result of $\alpha=1/2$, but holds the same form as that of $\alpha=1$. Besides, our work is a generalization on the previous work about the generalized Jarzynski equality based on the Feynman-Kac formula in the inhomogeneous diffusion process \cite{ge2008generalized}. The diffusion process considered there is under a specific stochastic interpretation: the Fokker-Planck equation is under Ito's interpretation.

\section{Detailed derivation on the main result}
\label{section 3}
In this section, we demonstrate how to derive Eq.~(\ref{generalized FT}) by first introducing the reverse process. For the dynamics described by Eq.~(\ref{Fokker-Planck}), we take the reverse process governed by the reverse protocol $\lambda^{R}(t)=\lambda^{F}(-t)$ and the following transformation $\mathcal{T}$ \cite{arenas2012hidden,PhysRevE.85.041122}:
\begin{align}
\label{transformation}
&t\rightarrow -t,\,\alpha\rightarrow(1-\alpha),\,f\rightarrow f-(1-2\alpha)g^{'}g.
\end{align}
The idea on the construction of this transformation is as follows. First, we let $t\rightarrow -t$, and then we need $\alpha\rightarrow(1-\alpha)$ to ensure our observation on the trajectories for the forward and the reverse processes is at the same series of points on the time axis. Second, after replacing $\alpha$ by $1-\alpha$, the equilibrium state distribution for Eq.~(\ref{Fokker-Planck}) is changed. To guarantee that the reverse process subjects to the $\alpha$-interpretation Fokker-Planck equation with a reverse current (only with $t\rightarrow -t$ in Eq.~(\ref{Fokker-Planck})), we should also modify the drift term accordingly, which leads to the transformation $\mathcal{T}$. Thus, this transformation ensures that the forward and the reverse processes converge to an unique form of the equilibrium state distribution, and we will use this property in the following derivation. In a recent experiment realizing Stratonovich-to-Ito transition \cite{pesce2013stratonovich}, the drift part added with the term $\alpha g^{'}g$ ($\alpha\in[0,0.5]$) can be implemented. Therefore, the the reverse process given by the transformation $\mathcal{T}$ is achievable experimentally.

We next give the detailed derivation on Eq.~(\ref{generalized FT}) through calculating the ratio between the transition probabilities of the forward dynamical process and the corresponding reverse process. To guarantee that the forward and the reverse processes are described by the same set of sampling points, the fluctuation theorem for the overdamped Langevin dynamics with multiplicative noise was usually based on the path integral under Stratonovich's interpretation \cite{chernyak2006path,lahiri2013fluctuation}. Our derivation on the fluctuation theorem here is for the general stochastic interpretation by applying a path integral formulation provided in the appendix \ref{appendix 1}. It follows the path integral formulation recently developed for the overdamped Langevin dynamics without the control parameter \cite{tang2014summing}, which has significant differences compared with the previous path integral formulas \cite{hunt1981path,lau2007state,arenas2010functional} and can lead to correct transition probabilities in examples for the general stochastic interpretation.

According to the path integral formulation derived in the appendix \ref{appendix 1}, the conditional probability of observing a specific trajectory $\{x(t)|t_{0}\leq t\leq t_{N}\}$ in the forward process is:
\begin{align}
&P^{F}(x_{N}t_{N}|x_{0}t_{0})
=\exp\Big\{-\int_{t_{0}}^{t_{N}}\mathcal{S}_{+}[x(t),\dot{x}(t);\lambda^{F}(t)]dt\Big\},
\end{align}
with $\mathcal{S}_{+}[x(t),\dot{x}(t);\lambda^{F}(t)]=[\dot{x}-f-(\alpha-1/2)g^{'}g]^{2}/(2g^{2})+(g/2)[f/g+(\alpha-1/2)g^{'}]^{'}-(\partial\ln g/\partial\lambda)\dot{\lambda}^{F}/2$. Here, we do not write down the measure and will add them when doing the ensemble average over trajectories.

Next, let the corresponding reverse trajectory be $\{x^{\dag}(t)|x^{\dag}(t)=x(-t)\}$. Then, the conditional probability of this reverse trajectory is:
\begin{align}
&P^{R}(x^{\dag}_{N}t_{N}^{\dag}|x_{0}^{\dag}t_{0}^{\dag})
=\exp\Big\{-\int_{t_{0}}^{t_{N}}\mathcal{S}_{-}[x(t),\dot{x}(t);\lambda^{F}(t)]dt\Big\},
\end{align}
where $\mathcal{S}_{-}[x(t),\dot{x}(t);\lambda^{F}(t)]\doteq\mathcal{T}\mathcal{S}_{+}[x(t),\dot{x}(t);\lambda^{F}(t)]$.
Under the transformation $\mathcal{T}$, the term $g[f/g+(\alpha-1/2)g^{'}]^{'}/2$ is invariant. We thus have
$S_{+}(x,\dot{x};\lambda^{F})-S_{-}(x,\dot{x};\lambda^{F})
=\dot{x}\partial_{x}\phi-(2\alpha-1)\dot{x}g^{'}/g-(\partial\ln g/\partial\lambda)\dot{\lambda}^{F}$. The ratio between the conditional probabilities is:
\begin{align}
\frac{P^{F}(x_{N} t_{N}| x_{0} t_{0})}{P^{R}(x^{\dag}_{N} t^{\dag}_{N}| x^{\dag}_{0}t^{\dag}_{0})}
&=\exp\Big\{-\int_{t_{0}}^{t_{N}}\dot{x}\Big[\partial_{x}\phi-(2\alpha-1)\frac{g^{'}}{g}\Big]dt
\notag\\&\quad+\int_{t_{0}}^{t_{N}}\frac{\partial\ln g}{\partial\lambda}\dot{\lambda}^{F}dt\Big\}.
\end{align}

The ratio between the unconditional probabilities is obtained by multiplying the initial distributions. When we choose $\rho_{eq}$ as the form of the initial distributions for both the forward and the reverse precesses, we have:
\begin{align}
\label{unconditional}
\frac{P^{F}(x)}{P^{R}(x^{\dag})}&=\frac{g[x_{N},\lambda(t_{N})]}{g[x_{0},\lambda(t_{0})]}\exp(\widetilde{W}^{F}).
\end{align}

To ensure the normalization condition for the probability, the measure for the initial distribution should be given according to $\rho_{eq}$. Thus, the measure for the forward process is:
\begin{align}
\label{measure1}
\int\mathcal{D}x&=\frac{dx_{0}}{Z_{\alpha}[\lambda(t_{0})]}\lim_{N\rightarrow\infty}\prod^{N}_{n=1}\frac{dx_{n}}{\sqrt{2\pi\tau }g[x_{n},\lambda(t_{n})]}.
\end{align}
For the reverse process, the corresponding measure can be written as follows:
\begin{align}
\label{measure2}
\int\mathcal{D}x^{\dag}&=\frac{dx_{0}^{\dag}}{Z_{\alpha}(\lambda(t_{0}^{\dag}))}\lim_{N\rightarrow\infty}\prod^{N}_{n=1}\frac{dx^{\dag}_{n}}{\sqrt{2\pi\tau }g[x_{n}^{\dag},\lambda(t_{n}^{\dag})]}
\notag\\&=\frac{dx_{N}}{Z_{\alpha}[\lambda(t_{N})]}\lim_{N\rightarrow\infty}\prod^{N-1}_{n=0}\frac{dx_{n}}{\sqrt{2\pi\tau }g[x_{n},\lambda(t_{n})]},
\end{align}
where we have used the conjugate relation for the forward and the reverse trajectories.

Let $\rho^{F}(\widetilde{W})$ denote the distribution of $\widetilde{W}$ values by a realization through the path integral for the forward process, and $\rho^{R}(\widetilde{W})$ for the reverse process. We define $\widetilde{W}^{R}\doteq\int_{t_{0}}^{t_{N}}\dot{\lambda}^{R}(\partial{V_{eq}}/\partial\lambda)dt$. Then, $\widetilde{W}$ is odd under the time-reversal in the sense that: $\widetilde{W}^{R}(x^{\dag})=-\widetilde{W}^{F}(x)$. Combing Eq.~(\ref{unconditional}), Eq.~(\ref{measure1}) and Eq.~(\ref{measure2}), we have
\begin{align}
\label{FT}
&\rho^{F}(\widetilde{W})
\notag\\&=\int\mathcal{D}xP^{F}(x)\delta(\widetilde{W}-\widetilde{W}^{F}(x))
\notag\\&=\exp(\widetilde{W})\frac{Z_{\alpha}[\lambda(t_{N})]}{Z_{\alpha}[\lambda(t_{0})]}\int\mathcal{D}x^{\dag}P^{R}(x^{\dag})\delta(\widetilde{W}+\widetilde{W}^{R}(x^{\dag}))
\notag\\&=\exp(\widetilde{W})\frac{Z_{\alpha}[\lambda(t_{N})]}{Z_{\alpha}[\lambda(t_{0})]}\rho^{R}(-\widetilde{W}),
\end{align}
which leads to Eq.~(\ref{generalized FT}).

\section{Conclusion}
\label{section 4}
We have obtained a series of non-equilibrium work relations as generalization of the Jarzynski equality, the detailed fluctuation theorem and the Hatano-Sasa's relation for the overdamped Langevin dynamics with multiplicative noise. Our result has demonstrated that, in the presence of multiplicative noise, the free energy change calculated by non-equilibrium work measurements using the Jarzynski equality remains valid only for the system with the Boltzmann-Gibbs distribution at the equilibrium state. For systems with general stochastic interpretations of multiplicative noise, the generalized Jarzynski equality has provided a connection between the non-equilibrium work measurements and the generalized partition function. The robustness of the Hatano-Sasa's relation with respect to the stochastic interpretation has also been shown. The non-equilibrium thermodynamical equalities derived here remain to be tested experimentally.

For the overdamped Langevin dynamics with multiplicative noise, recent works on a decomposition of the dynamical system \cite{ao2004potential,kwon2005structure,ao2008emerging} provide a constructive method to find the potential function leading to the Boltzmann-Gibbs distribution. This framework assigns a specific stochastic interpretation for the Langevin dynamics, which corresponds to anti-Ito's in one dimension and goes beyond the $\alpha$-interpretation when dimension is larger than one \cite{yuan2012beyond}. With this framework, how to generalize our derivation here to systems without detailed balance is an interesting topic to be explored.

\section*{Acknowledgments}
Stimulating discussions with Hao Ge, Alberto Imparato, David Cai, Hong Qian, Christopher Jarzynski, Andy Lau and Xiangjun Xing are gratefully acknowledged. We thank Zhu Xiaomei for the critical comments. This work is supported in part by the National 973 Project No.~2010CB529200 and by the Natural Science Foundation of China Projects No.~NSFC61073087 and No.~NSFC91029738. Ying Tang was partially supported by an Undergraduate Research Program in Zhiyuan College at Shanghai Jiao Tong University.

\section{Appendix \textrm{A}: The path integral framework}
\appendix
\setcounter{section}{1}
\label{appendix 1}
In this appendix, we provide the path integral formulation, which is used to derive the generalized fluctuation theorem in the main text. This formulation follows our previous work on the path integral construction for the Langevin dynamics without the external control parameter \cite{tang2014summing}.

For the Langevin equation Eq.~(\ref{Langevin1}) under the $\alpha$-interpretation, by modifying the drift term, we have its equivalent Langevin equation under the Stratonovich's interpretation \cite{gardiner2004handbook}:
\begin{align}
\label{Langevin2}
\dot{x}=f(x;\lambda)+\Big(\alpha-\frac{1}{2}\Big)g^{'}g(x;\lambda)+g(x;\lambda)\xi(t),
\end{align}
where the superscript prime denotes the derivative to $x$. The advantage of using this Stratonovich's form is that ordinary calculus rule can be simply applied \cite{arenas2012hidden}. Then, this equation can be transformed to be a Langevin equation with a additive noise by a change of variable $q=H(x;\lambda)$ with $H^{'}(x;\lambda)=1/g(x;\lambda)$ \cite{hunt1981path}:
\begin{align}
\label{Langevin4}
\dot{q}-h(q;\lambda)=\xi(t),
\end{align}
where we have introduced an auxiliary function:
\begin{align}
h(q;\lambda)&=\frac{f\big(H^{-1}(q);\lambda\big)}{g\big(H^{-1}(q);\lambda\big)}+\Big(\alpha-\frac{1}{2}\Big)g^{'}\big(H^{-1}(q);\lambda\big)
\notag\\&\quad+\frac{\partial H\big(H^{-1}(q);\lambda\big)}{\partial\lambda}\dot{\lambda}.
\end{align}

To get the transition probability for Eq.~(\ref{Langevin4}), we first discretize the time into $N$ segments: $t_{0}<t_{1}<\dots<t_{N-1}<t_{N}$ with $\tau= t_{n}-t_{n-1}$ small and let $q_{n}=q(t_{n})$, $\lambda_{n}=\lambda(t_{n})$. For the sake of consistency, as we have chosen the equivalent Stratonovich's form, the corresponding discretized Langevin equation needs the mid-point discretization:
\begin{align}
\label{Langevin_discret}
q_{n}-q_{n-1}-\frac{h(q_{n};\lambda_{n})+h(q_{n-1};\lambda_{n-1})}{2}\tau=W_{n}-W_{n-1},
\end{align}
where $W(t)$ is the Wiener process given by $dW(t)=\xi(t)dt$. Thus, the Jacobian for the variable transformation between $q(t)$ and $W(t)$ is:
\begin{align}
&J\approx\exp\Big[-\frac{\tau}{2}\sum_{n=1}^{N-1}\frac{\partial h(q_{n};\lambda_{n})}{\partial q_{n}}\Big].
\end{align}

Then, with the property of the Wiener process and the Chapman-Kolmogorov equation \cite{gardiner2004handbook}, the path integral formulation for Eq.~(\ref{Langevin4}) is obtained:
\begin{align}
\label{transitional probability1}
&P(q_{N} t_{N}| q_{0} t_{0})
\notag\\&=\int^{q_{N}}_{q_{0}}\mathcal{D}q\exp\Big\{-\int^{t_{N}}_{t_{0}}\Big[\frac{1}{2}(\dot{q}-h)^{2}+\frac{1}{2}\frac{\partial h}{\partial q}\Big]dt\Big\},
\end{align}
where $\int^{q_{N}}_{q_{0}}\mathcal{D}q\doteq\lim_{N\rightarrow\infty}\frac{1}{\sqrt{2\pi\tau}}\prod^{N-1}_{n=1}\int\frac{dq_{n}}{\sqrt{2\pi\tau}}$. The integral of the action function on the exponent obeys ordinary calculus due to the mid-point discretization and the last term comes from the Jacobian.

By changing the variable reversely: $x=H^{-1}(q;\lambda)$ with $\partial x/\partial q=g(x;\lambda)$, we get the path integral for Eq.~(\ref{Langevin1}) under the $\alpha$-interpretation:
\begin{widetext}
\begin{align}
\label{result1}
&P(x_{N} t_{N}| x_{0} t_{0})=\int^{x_{N}}_{x_{0}}\mathcal{D}x\exp\Big\{-\int^{t_{N}}_{t_{0}}\Big[\frac{1}{2g^{2}}\Big(\dot{x}-f
-\Big(\alpha-\frac{1}{2}\Big)g^{'}g\Big)^{2}+\frac{g}{2}\Big(\frac{f}{g}+\Big(\alpha-\frac{1}{2}\Big)g^{'}\Big)^{'}-\frac{1}{2}\frac{\partial\ln g}{\partial\lambda}\dot{\lambda}\Big]dt\Big\},
\end{align}
\end{widetext}
where the measure is:
\begin{align}
\int^{x_{N}}_{x_{0}}\mathcal{D}x&\doteq\lim_{N\rightarrow\infty}\frac{1}{\sqrt{2\pi\tau}g(x_{N};\lambda_{N})}\prod^{N-1}_{n=1}\int\frac{dx_{n}}{\sqrt{2\pi\tau}g(x_{n};\lambda_{n})}.
\end{align}
Though the Jacobian term comes from the measure transformation and does not belong to the conventional action part, it is usually included in the action function for applications.

The path integral formulation derived in \cite{tang2014summing} has significant differences compared with the previous path integral formulas \cite{hunt1981path,lau2007state,arenas2010functional}. It shows that the path integral formulation for the overdamped Langevin equation with multiplicative noise is not unique but $\alpha$-dependent, and can generate the $\alpha$-interpretation Fokker-Planck equation \cite{shi2012relation}. It also leads to transition probabilities obeying the conservation law for general stochastic interpretations in examples.

\bibliography{bib}
\end{document}